# Successful growth of low carrier density α-In$_2$Se$_3$ single crystals using Se-flux in a modified Bridgman furnace

Soumi Mondal[1], Sreekant Anil[2], Saurav Islam[1], Yingdong Guan[1], Sai Venkata Gayathri Ayyagari[2], Aaron Pearre[1], Sandra Santhosh[1], Nasim Alem[2], Nitin Samarth[1], Zhiqiang Mao[1,2a]

[1]Department of Physics, The Pennsylvania State University, University Park, Pennsylvania 16802, USA

[2]Department of Materials Science and Engineering, The Pennsylvania State University, University Park, PA 16802, USA

## Abstract

Indium selenide (In$_2$Se$_3$) has garnered significant attention for its intriguing properties and applications in batteries, solar cells, photodetectors and ferroelectric devices. However, the controlled synthesis of single phase α −In$_2$Se$_3$ remains challenging owing to its complex phase diagram, presence of multiple polymorphs and the high volatility of selenium that induces non-stoichiometry and unintentional carrier doping. For ferroelectric α −In$_2$Se$_3$, minimizing the carrier density is essential because leakage current can obscure polarization switching. Here, we report the growth of α −In$_2$Se$_3$ single crystals using a unique approach, the Se-flux assisted modified vertical Bridgman technique combined with liquid encapsulation under high pressure. This approach creates a high-pressure, Se-rich environment that effectively minimizes Se-vaporization. Structural and compositional analysis using X-ray diffraction, transmission electron microscopy and energy-dispersive X-ray spectroscopy confirm the formation of pure α −In$_2$Se$_3$ single crystals with 3R stacking. Furthermore, the crystals exhibit remarkably low carrier density of 1.5-3.2×10$^{16}$ cm$^{-3}$ at 300K—the lowest reported to date, reflecting a significant suppression of Se-vacancies relative to the conventional Bridgman or melt-grown crystals. Through transport and ARPES measurements on different batches of crystals, we also demonstrate that the amount of Se-flux plays a crucial role in controlling Se-vacancies. Our results thus establish this modified Bridgman method as an effective strategy for synthesizing large α −In$_2$Se$_3$ single crystals with reduced intrinsic defects. This technique can be broadly applied to grow other volatile chalcogenides with reduced defects and controlled stoichiometry.

[a] Author to whom correspondence should be addressed: zim1@psu.edu

**Introduction**

Indium selenide ($In_2Se_3$) has emerged as a technologically important material owing to its remarkable electrical, optical as well as thermoelectric properties and potential applications in ionic batteries, solar cells, photodiodes, phase-change memory and other optoelectronic devices [1–8]. Among its polymorphs, the $\alpha-In_2Se_3$ has attracted particular attention, due to the observation of room-temperature ferroelectricity in the ultrathin flakes[9–15]. More recently, $\alpha-In_2Se_3$ has been found to exhibit bulk photovoltaic effect (BPVE) and photo-ferroelectric response, further extending its potential for next generation energy-efficient photodetectors and terahertz (THz) applications[16–18].

Despite its promise, controlled synthesis of $\alpha-In_2Se_3$ remains non-trivial. $In_2Se_3$ exhibits rich polymorphism – $\alpha, \beta, \gamma, \delta$ phases with temperature-dependent phase transitions ($\alpha \to \beta$ at $\sim 200^\circ C$, $\beta \to \gamma \to \delta$ at higher temperatures, as shown in **Fig. 1b**) frequently resulting in mixed-phase crystals[19–24]. Moreover, the $\alpha-In_2Se_3$ itself comprises two stacking variants – 2H ($P6_3mc$) and 3R ($R3m$)[19,20], often coexisting in one sample[13]. The complexity of In-Se phase diagram, which includes other stable stoichiometries such as InSe, $In_4Se_3$, and $In_6Se_7$ [23], along with selenium's high vapor pressure that induces Se loss and vacancies, further complicates the growth. As a result, the $In_2Se_3$ crystals grown using conventional methods such as Bridgman or melt growth often exhibit phase mixing and high defect densities [13,19–22,24,25]. Although ferroelectricity in $\alpha-In_2Se_3$ has been inferred from piezoresponse force microscopy (PFM) on thin-flakes and tunnel junctions, reliable polarization-electric field (P-E) loops measurements remain elusive due to leakage current arising from typically high carrier concentrations.

Until now, several approaches have been explored to synthesize $\alpha-In_2Se_3$ single crystals, including melt growth[19,21,22,26], chemical vapor transport (CVT)[8,19,27] and, most commonly, the Bridgman and temperature-gradient methods [28–33]. However, CVT and Bridgman grown crystals typically exhibit high room temperature electron densities in the range of $5\times10^{17} - 2\times10^{18}$ cm$^{-3}$ [29–32], indicating a significant concentration of Se vacancies, which limits the material's electronic and ferroelectric performance. Although long-term post-growth annealing was reported to be effective in reducing the carrier density down to $3\text{-}7\times10^{16}$ cm$^{-3}$ at 300K [24,26,34–36], these studies were performed on polycrystals or "almost single crystals" and lacked proper structural characterization, implying those reported low carrier density may not be intrinsic. These challenges highlight the need for single-phase $\alpha-In_2Se_3$ single crystals with controlled stoichiometry and significantly reduced carrier density, without compromising the crystalline quality or phase purity. More recently, high pressure and high temperature (HPHT) method[37] and growth from saturated solutions in Se vapor[38] have been reported to achieve controlled growth of $\alpha-In_2Se_3$. However, the carrier density was not reported for those samples.

In this work, we introduce a unique crystal growth method, the Se-flux assisted modified vertical Bridgman method to grow pure $\alpha-In_2Se_3$ single crystals. This modified Bridgman furnace uses a double-crucible geometry as shown in **Fig. 1a**, enabling continuous feeding of source material from the upper crucible to the lower growth crucible, and hence also called double crucible vertical Bridgman (DCVB). In addition, it supports liquid encapsulation and high-pressure growth, both of which are advantageous for volatile chalcogenide growth as it can suppress vaporization of the volatile element. Our group has previously demonstrated the successful synthesis of high quality $Bi_2Se_3$ single crystals with reduced defect using this novel method[39].

Although the Bridgman growth of $In_2Se_3$ using small amounts of excess In (6 mol%)[29] and excess Se (2-3 mol % excess)[40] has been reported, no prior work utilized the Se-flux environment to grow $In_2Se_3$ single crystals so far. The In-Se phase diagram **(Fig. 1b)** indicates that stoichiometric $In_2Se_3$ crystals can probably be grown from melts containing highly excessive Se near the eutectic concentration, though using significant Se exceeding the eutectic concentration will drive the formation of $In_2Se_3$–Se mixture. Motivated by this idea, we attempted Se-flux growth with a significant Se-flux of 20% and 29.3% excessive relative to stoichiometric Se, marked by two red dashed lines below the eutectic point **in Fig. 1b**. To prevent this excess Se from vaporizing, boron oxide ($B_2O_3$) was chosen as the liquid encapsulant. Due to its chemical inertness, low density and relatively low melting point (~450°C) than the III-VI semiconductor, $B_2O_3$ forms a protective liquid layer over the melt, effectively minimizing the vaporization of Se under high pressure. Although the continuous feeding mode could not be utilized for $In_2Se_3$ growth due to its high volatility, we successfully synthesized single crystals of $\alpha-In_2Se_3$ with significantly reduced Se vacancies using this modified Bridgman technique.

Structural and compositional characterization from X-ray diffraction (XRD), transmission electron microscopy (TEM) and energy-dispersive X-ray spectroscopy (EDXS) confirm the synthesis of single-phase $\alpha-In_2Se_3$ single crystal with 3R stacking. Most notably, our $\alpha-In_2Se_3$ single crystals exhibit a remarkably low carrier density of $1.5\times10^{16}$ cm$^{-3}$ at 300K, which further decreases to $4\times10^{15}$ cm$^{-3}$ at 10K. To the best of our knowledge, these values represent the lowest carrier densities for $\alpha-In_2Se_3$ single crystals reported to date and are 1-2 order of magnitudes lower than the crystals grown by conventional Bridgman methods ($5\times10^{17} - 2\times10^{18}$ cm$^{-3}$ at 300K)[29–32]. This substantial reduction in carrier density correlates with the reduction in Se vacancies achieved by using Se-flux during the growth. These results establish the Se-flux-assisted modified Bridgman method, combined with liquid encapsulation under high pressure, as a promising pathway for growing large, single-phase $\alpha-In_2Se_3$ single crystals with much lower defect concentration.

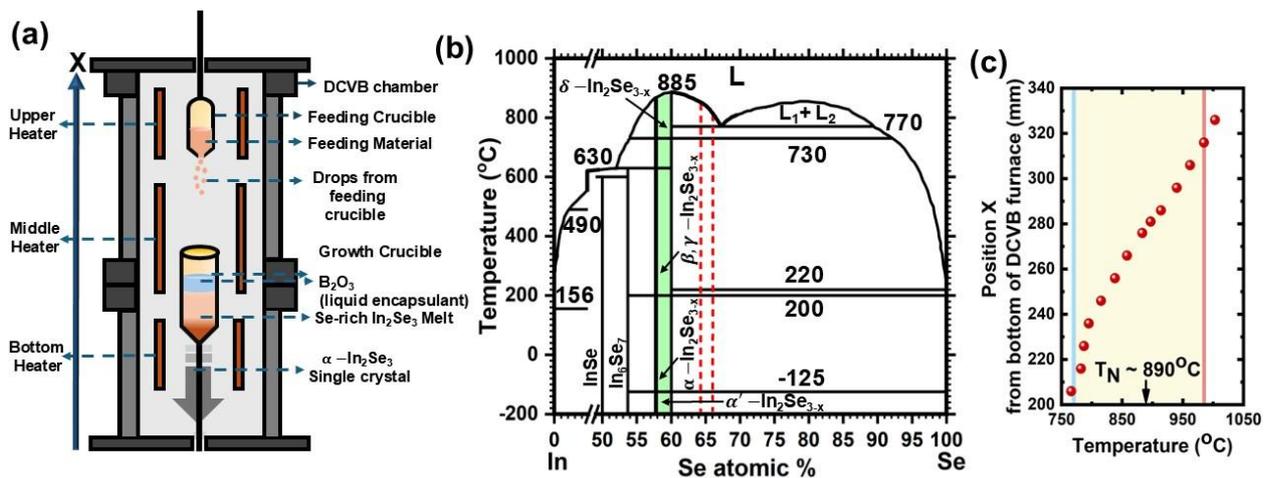

**Figure 1.** (a) Schematic of modified Bridgman Furnace with double-crucible geometry. (b) Schematic of the In-Se phase diagram obtained from previous reports[23,41,42]. The green region corresponds $In_2Se_{3-x}$ ($In_2Se_3$ with Se vacancies) and the two red dashed lines indicate the Se concentrations used for growing $In_2Se_3$. (c) The temperature profile used for the growth of $In_2Se_3$ crystal. The start and the end temperatures of the translation are marked by the red and blue lines respectively, while the yellow background shows the entire temperature range over which the crystal growth occurred. $T_N$ refers to the melting point of $In_2Se_3$.

**Methods**

$\alpha-In_2Se_3$ crystals were grown by employing a combination of Bridgman and flux-growth methods in presence of liquid encapsulation and high pressure (up to 10 atm) in a modified Bridgman furnace. Combination of liquid encapsulation and high pressure helps to suppress the vaporization of Se and allow better control of the melt composition during crystal growth. As noted above, this furnace uses the concept of double crucible (**Fig. 1a**), where the lower growth crucible contains the source materials and the upper crucible supplies feeding materials to the growth crucible, as shown in **Fig. 1a**. In this modified technique, the crucible loaded with the source material can be directly placed inside the high-pressure chamber, without sealing in a quartz tube. A proper amount of $B_2O_3$ was added to the crucible, which serves as liquid encapsulation during the growth.

Initially, we attempted using the feeding crucible to continuously supply $In_2Se_3$ to the lower growth crucible, since this approach may enable the traveling solvent growth as demonstrated in the DCVB growth of $Bi_2Se_3$[39] and thus better control crystal stoichiometry. However, due to high volatile nature of $In_2Se_3$, significant vaporization occurred, leading to deposition on the upper walls and sides of the chamber instead of dripping down into the growth crucible. As a result, this feeding approach proved unsuitable for $In_2Se_3$. Nevertheless, by adopting a Se-flux-assisted liquid encapsulation growth strategy under high pressure – without continuous feeding – we successfully synthesized $\alpha-In_2Se_3$ single crystals. As seen from the In-Se phase diagram in **Fig. 1b**, Se-flux growth can possibly favor the growth of $In_2Se_3$ crystals with less Se vacancies. Thus, we chose two different Se-flux concentrations (marked by red lines in **Fig. 1b**) close to the eutectic point to perform the growth.

High-purity In and Se powders were weighed out in a glove box under an inert atmosphere in a molar ratio of 34:66 (~2:3.88) corresponding to 29.3 % excess Se, which acted as a flux (Batch-1). For Batch-2, In and Se powders were loaded in a molar ratio of 40:72 (~ 2:3.6), corresponding to 20% Se flux. These elemental powders were thoroughly mixed and loaded in a tip-shaped alumina crucible. To prevent the vaporization of Se during heating, ~1.5g of $B_2O_3$ was added on top of the In-Se mixture, acting as a liquid encapsulant. The alumina crucible was then carefully mounted on the bottom shaft of the furnace, serving as the growth crucible.

After mounting the growth crucible, the chamber was evacuated to a vacuum level of ~$10^{-4}$ torr and then filled with Ar gas up to a pressure of 10 atm. The power of the three carbon heaters of the furnace (**Fig. 1a**) was adjusted to obtain the desired temperature gradient of ~25°C/cm suitable for the growth. Using the required input power, the temperature at the initial position of the growth crucible (close to the center of the middle zone) was set to 980°C which is above the melting point of In$_2$Se$_3$ (890°C), while the bottom zone was maintained at 770°C, creating the desired temperature gradient. The system was held under these conditions for 24hrs to allow homogenization of the melt. After this, crystal growth was initiated by translating the growth crucible downward from the high temperature position (980°C) to the low temperature position (770°C) at a controlled rate of 0.9 – 1.0 mm/hr. The temperature profile used for this growth is presented in **Fig. 1c**. Upon completion of the translation process, the furnace was slowly cooled down to room temperature over a period of 48hrs by gradually ramping down the heater power.

After cooling, the chamber was opened and only minimal Se deposition was observed on the growth wall chamber, confirming the effectiveness of the use of B$_2$O$_3$ liquid encapsulation and high pressure in minimizing Se vaporization. Finally, the alumina crucible was carefully broken to obtain the crystal ingot. **Figure 2a** shows the image of the Batch-1 $\alpha-$In$_2$Se$_3$ crystal grown by this new method.

XRD was used to identify the phase of the grown crystals, whereas crystal composition was analyzed using EDXS. The stacking arrangement in our $\alpha-$In$_2$Se$_3$ crystal was determined by TEM. For both TEM and STEM analyses, the lamella was prepared along the zone axis [$\bar{2}$110] using focused ion beam (FIB) techniques. Thinning for electron transparency was accomplished with a Thermo Fisher Scientific Helios 660 NanoLab Dual-Beam FIB. Aberration-corrected STEM imaging was conducted on a Thermo Fisher Scientific Titan G2 microscope operated at 300 kV. A bandpass filter was applied to the acquired ADF-STEM images using the bandpass filtering function in GATAN Digital Micrograph. Selected-area electron diffraction (SAED) patterns and TEM images were obtained using the Thermo Fisher Scientific Talos F200X microscope at an accelerating voltage of 200 kV.

The electrical transport measurements of $\alpha-$In$_2$Se$_3$ crystals were done using the four-probe method in Quantum Design Physical Property Measurement System (PPMS). Angle-resolved photoemission spectroscopy (ARPES) measurements were carried out at 300K using the 21eV Helium I$\alpha$ spectral line from a helium plasma lamp, isolated using a monochromator and the emitted photoelectrons were detected using Scienta-Omicron DA 30L analyzer having a spectral resolution of 6 meV. Prior to measurements, all the samples were cleaved inside the chamber at a base pressure of 5×10$^{-9}$ mbar to obtain a pristine surface.

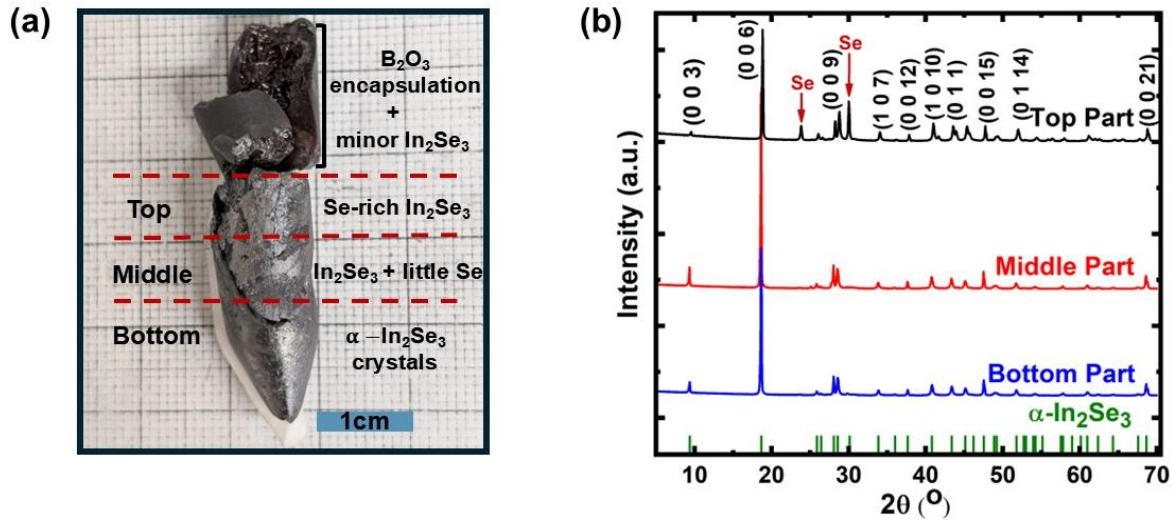

**Figure 2.** (a) Image of the crystal ingot grown using the DCVB Furnace without feeding (batch-1). The ingot is divided into four regions: $B_2O_3$ with some minor $In_2Se_3$, Top which corresponds to Se-rich $In_2Se_3$, Middle part which is $In_2Se_3$ with little Se and Bottom part which is pure $\alpha-In_2Se_3$ single crystal. (b) Powder XRD of $\alpha-In_2Se_3$ from the three different regions: Top, Middle and Bottom. Top part shows a very strong additional peak corresponding to elemental Se, which is very weak in the XRD from middle part and absent for the bottom region. Our XRD pattern matches with the simulated pattern for $\alpha-In_2Se_3$ (in green) obtained from ICDD database (PDF-5+ 2026, 04-027-0925)

### Characterization of the Batch-1 $In_2Se_3$ crystal

The top brownish layer above the grown crystal ingot **(Fig. 2a)** was identified to consist primarily of $B_2O_3$ with some minor incorporation of $In_2Se_3$, which accounts for the coloration. Minimal Se contamination was observed on the growth chamber walls which indicates that $B_2O_3$ served as an effective encapsulant, though with slight mixing of $In_2Se_3$. To investigate the composition and structural phase of the grown crystal, EDXS and XRD analysis have been performed. Based on these characterizations, the crystal is divided into three regions — Top, Middle and Bottom.

Powder XRD from the three regions of the crystal is shown in **Fig. 2b**. No secondary phases were detected in the bottom region of the crystal, and all the diffraction peaks matched with those of $\alpha-In_2Se_3$, confirming the growth of pure $\alpha-In_2Se_3$. In contrast, the top region exhibited an additional strong diffraction peak corresponding to elemental Se, while the middle region displayed a very weak peak of Se. These results from XRD agree with the EDXS results presented in **Fig. 3**.

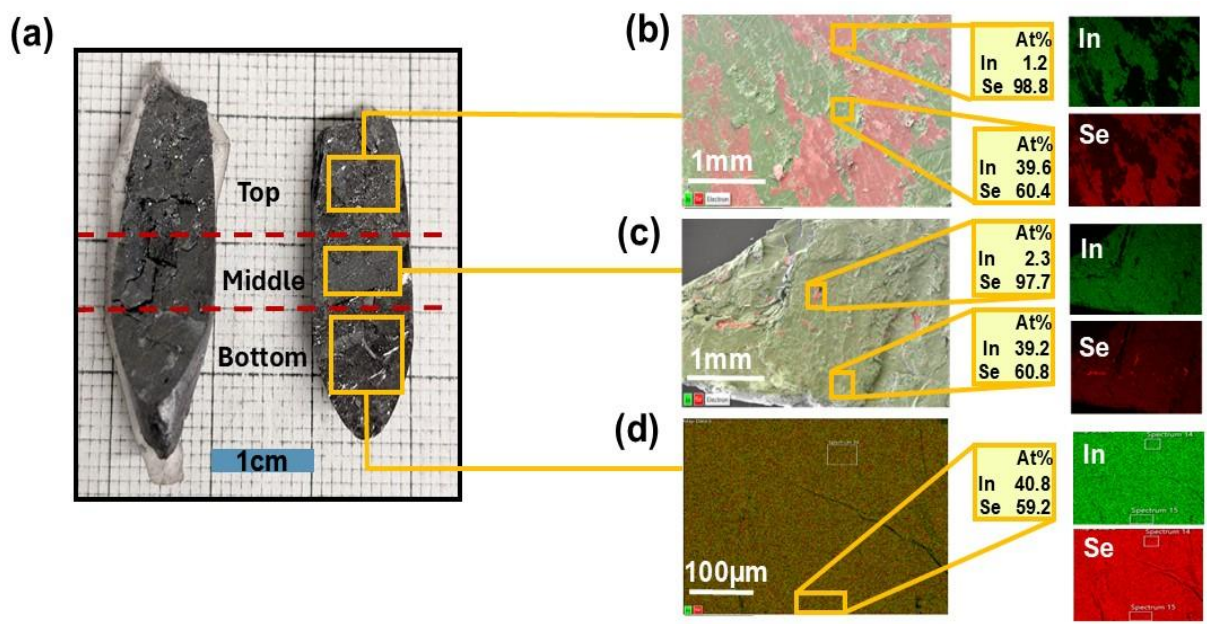

**Figure 3**. (a) Image of the cleaved crystals of Batch-1 $\alpha-In_2Se_3$. (b)-(d) EDXS mapping of single crystal piece cleaved from three different regions Top (b), Middle (c) and Bottom (d). Green regions correspond to $In_2Se_3$ and the red regions to elemental Se. The top region of the crystal has rich Se, while the middle part contains very little Se. Pure $In_2Se_3$ with homogenous distribution of In and Se is obtained from the bottom region of the ingot.

The top region located just below the $B_2O_3$ is characterized by more significant precipitates of Se, as evident from the EDXS mapping of the crystal piece cleaved from the top part **(Fig. 3b)**. The green regions correspond to $In_2Se_3$, and the red regions denote presence of pure Se. The middle region showed only minor Se precipitates **(Fig. 3c)**, suggesting the formation of almost pure $In_2Se_3$. The crystal cleaved from the bottom region reveals a homogeneous distribution of In and Se, confirming the presence of pure $In_2Se_3$ crystals **(Fig. 3d)**. As $In_2Se_3$ starts to crystallize from the melt, the relative proportion of excess Se increases in the remaining melt driving the composition of the melt towards the $In_2Se_3$-Se eutectic point which leads to the formation of $In_2Se_3$–Se mixture in the top region.

**Fig. 4a** shows the image of a crystal piece cleaved from the bottom region. The single-crystalline nature of the crystal is further confirmed by the strong and prominent (00l) reflections in **Fig. 4b**. Combined XRD and EDS analyses demonstrate that the use of Se flux together with liquid encapsulation under high pressure effectively facilitates the growth of large $\alpha-In_2Se_3$ single crystals.

Additionally, TEM was performed to determine whether the grown $\alpha-In_2Se_3$ has 3R or 2H stacking. **Figure 4c** compares the crystal structures of 3R and 2H $\alpha-In_2Se_3$. Both the crystal structures display a van der Waals (vdW) layered framework with each slab comprising covalently bonded Se-In-Se-In-Se quintuple layer stacked along the c-axis, while adjacent quintuple-layer slabs are coupled together by weak vdW forces. The difference between the 3R and 2H phase of $\alpha-In_2Se_3$ lies in the stacking of these layers[19,20,23,43]. The 3R $\alpha-In_2Se_3$ belongs to space group R3m with rhombohedral structure characterized by three-layer periodicity where successive monolayers show a small translation along the ab plane, resulting in ABC stacking **(Fig. 4c)**. In contrast, the 2H $\alpha-In_2Se_3$ has a hexagonal structure with space group P6$_3$mc and exhibit AB stacking with two-layer periodicity **(Fig. 4c).**

A representative cross-sectional selected-area electron diffraction (SAED) pattern of our crystal (**Fig. 4d**) corresponds to the [$\bar{2}$110] zone axis of 3R $\alpha-In_2Se_3$ structure (simulated electron diffraction pattern given in **Fig.S1**, in the supplementary materials). Moreover, the atomically resolved scanning transmission electron microscopy (STEM) images presented in **Fig. 4e and 4f** reveal a clear ABC stacking sequence, further confirming the 3R polytype. However, we also observe some additional reflections in the SAED pattern acquired from slightly different region of the crystal, as shown in **Fig. S2** in supplementary material, which likely originate from twin domains.

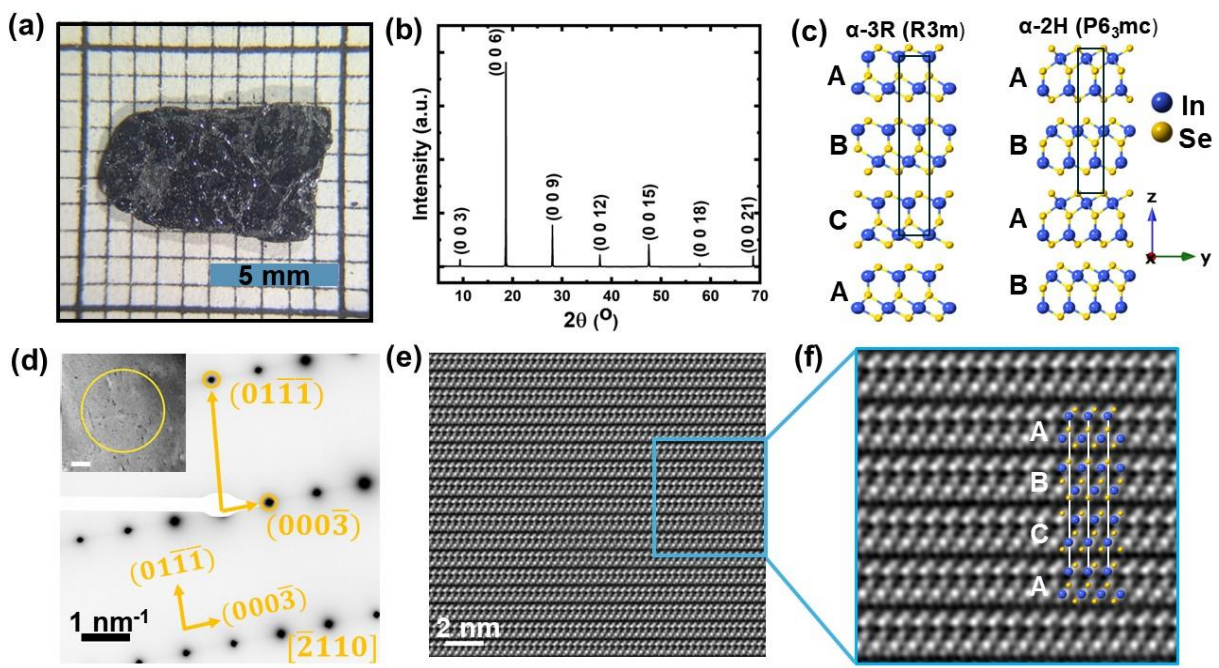

Figure 4. (a) Image of a cleaved Batch-1 crystal piece from the bottom region of the ingot. (b) XRD pattern of the crystal piece confirming the growth of $\alpha-In_2Se_3$ single crystal. (c) Crystal structures of the 2H and 3R $\alpha-In_2Se_3$ (d) SAED pattern of Batch-1 $\alpha-In_2Se_3$ crystal along the [$\bar{2}$110] zone with the real-space TEM image shown as inset (inset scale bar: 200nm) (e) ADF-STEM image of the crystal piece showing ABC stacking sequence, confirming 3R phase in Batch-1 $\alpha-In_2Se_3$. (f) A zoomed-in region in which the atomic model of 3R $\alpha-In_2Se_3$ projected along the direction [$\bar{2}$110] is overlaid. .

**Electronic property of Batch-1 α−In₂Se₃ crystal**

α − In₂Se₃ can exhibit a variety of defects such as cationic vacancy $V_{In}$, interstitial In atoms ($In_i$) and anionic selenium vacancy ($V_{Se}$)[25], however, undoped In₂Se₃ crystals are consistently observed to be n-type reflecting that Se vacancies are the most dominant defects in this compound[44]. Since the presence of Se vacancies act as electron donor, it is expected that reduction in the concentration of these vacancies will lower the carrier concentration ($n_e$), improving the crystal quality. Previous studies have reported that α −In₂Se₃ crystals, grown by Bridgman method, are electron doped with in-plane electron conductivity (σ) in range of 20-68.9 $\Omega^{-1}cm^{-1}$ and carrier density ranging from 5×10¹⁷ – 2×10¹⁸ cm⁻³ at 300K[29–32]. The α −In₂Se₃ single crystal grown with indium excess exhibited metallic like behavior with σ and $n_e$ increasing with decrease in temperature[29].

The influence of post-growth annealing on the electrical properties of α −In₂Se₃ was investigated several years ago as mentioned earlier[24,34,35]. Crystals synthesized by direct fusion of elements and subsequently subjected to prolonged annealing and slow cooling exhibited lower electron density of approximately 3-7×10¹⁶ cm⁻³, accompanied by low in-plane conductivity $\sigma_{in-plane}$ ~ 0.1 $\Omega^{-1}cm^{-1}$ at 300K whereas crystals quenched in liquid nitrogen showed higher electron densities of 1.4-2×10¹⁹ cm⁻³ and $\sigma_{in-plane}$ of around 260 $\Omega^{-1}cm^{-1}$ [24,34–36]. Furthermore, the quenched samples displayed an increasing σ upon cooling, whereas the annealed samples exhibited the opposite trend. These findings demonstrated that annealing the samples promoted defect annihilation, thereby leading to the expected decrease in the electron density. However, it is important to note that the reported measurements were performed on either polycrystalline or "almost single crystals" specimens[26], with no accompanying characterizations like XRD or EDXS. The absence of characterization results limits a clear assessment of the crystalline quality and defects in those samples and implies that the reported carrier density of the order of 10¹⁶ cm⁻³ is yet to verified.

To assess the crystal quality of our grown $\alpha-In_2Se_3$ single crystal, particularly the concentration of Se vacancies, electronic transport measurements have been carried out using a standard four-probe measurement technique. The in-plane electrical resistivity ($\rho_{xx}$) measured across the temperature range from 2.5K to 300K, shows an increase in $\rho_{xx}$ value upon cooling, suggesting insulating behavior. As shown in **Fig. 5a**, the measured $\rho_{xx}$ value is found to be 20 Ωcm at 300K, which is significantly larger than the values reported for $\alpha-In_2Se_3$ Bridgman-grown crystals (0.015-0.05 Ωcm at 300K)[28–32] and for the quenched samples ($3\times10^{-3}$ Ωcm)[24,34–36]. This value is also larger than the resistivity obtained for annealed samples (~10 Ωcm)[24,34–36]. The Hall resistivity ($\rho_{xy}$) versus magnetic field curves **(Fig. 5b)** exhibited a negative slope, confirming electrons as the dominant charge carriers in the sample, consistent with earlier reports.

The carrier concentration ($n_e$) values extracted from magnetic field-dependent Hall resistivity ($\rho_{xy}$) measurements at various temperatures are plotted as shown in **Fig. 5c.** At 300K, $n_e$ was determined to be $1.5\times10^{16}$ cm$^{-3}$, further dropping to a value of $4\times10^{15}$ cm$^{-3}$ at 10 K which is the lowest carrier density that has been reported till date for $\alpha-In_2Se_3$. For comparison, the previously reported carrier densities at 25K were around $5\times10^{16}$ cm$^{-3}$ for the annealed samples[34,35] and $\sim 10^{17}$ cm$^{-3}$ for crystals grown by the conventional Bridgman method[29–32]. Similar measurements performed on additional cleaved pieces of the same crystal also yielded $n_e$ in the range of 1.5-$3.2\times10^{16}$ cm$^{-3}$ at 300K, confirming the lowest carrier density in our grown α-In$_2$Se$_3$. These exceptionally low carrier densities, combined with high resistivity observed across multiple samples, provide strong evidence that Se-flux method produced α-In$_2$Se$_3$ single crystals with least density of Se vacancies reported so far. Despite our samples showing insulating behavior, we were unable to obtain reliable P-E hysteresis loops because of leakage currents, indicating that the crystals are still not sufficiently insulating for ferroelectric polarization measurements.

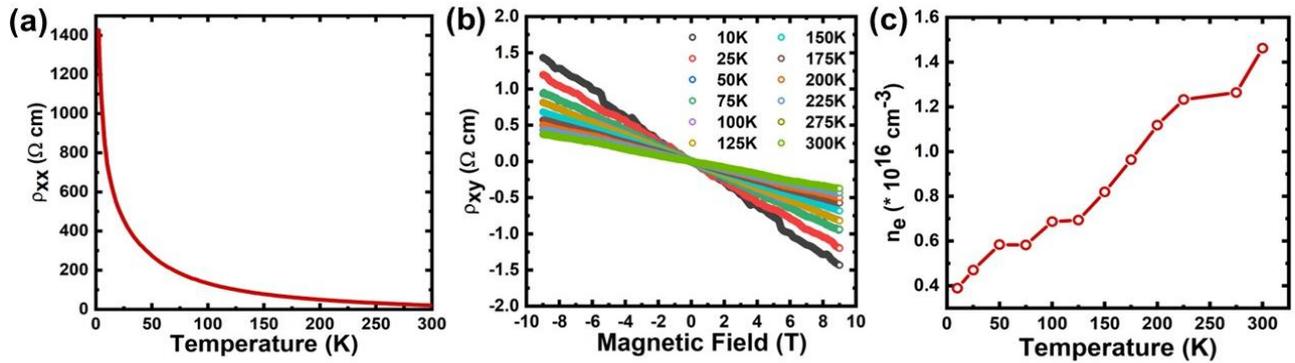

Figure 5. (a) Temperature dependent in-plane resistivity ($\rho_{xx}$) of $\alpha-In_2Se_3$ crystal cleaved from the bottom region. (b) Magnetic field-dependent Hall resistivity ($\rho_{xy}$) showing negative slope, confirming electrons to be the main carriers in the DCVB grown $\alpha-In_2Se_3$. (c) Temperature dependence of Carrier density ($n_e$) extracted from the Hall resistivity data

### Characterization and transport properties of Batch-2 In$_2$Se$_3$ crystals

The Batch-2 In$_2$Se$_3$ crystals were grown with 20% excess Se as mentioned in the methods section. The as-grown Batch-2 In$_2$Se$_3$ crystal ingot is shown in **Fig. 6a**. The ingot was divided into three regions — A, B and C for characterization similar to Batch-1 In$_2$Se$_3$ crystals. **Figure 6b** summarizes the EDXS composition analysis from all three regions. Unlike Batch-1 crystal which had a Se-rich top region, no such Se-enrichment was observed in Batch-2 crystal; instead, region A contained a mixed phase of In$_2$Se$_3$ and InSe. In contrast, the EDXS composition analysis of the crystal pieces cleaved from regions B and C confirms the formation of pure In$_2$Se$_3$ with a homogeneous composition. Furthermore, XRD scans of crystal pieces from regions B and C **(Fig. 6c)** show strong and sharp (00l) reflections confirming the growth of pure $\alpha-In_2Se_3$ single crystals.

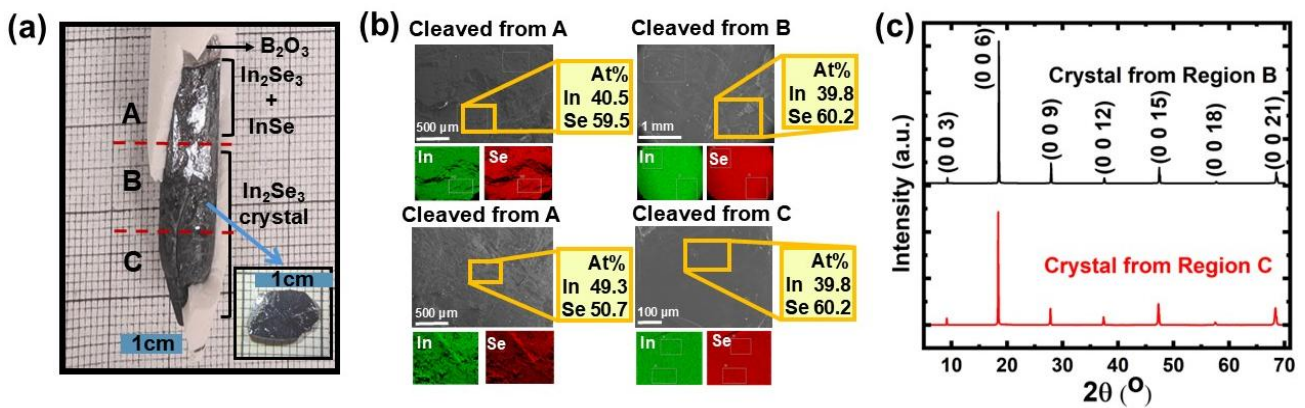

Figure 6. (a) Image of the Batch-2 In$_2$Se$_3$ crystal grown using with 20% excess selenium as flux. (b) EDXS elemental mapping and composition analysis of crystals from three different regions – A, B and C. Region A exhibits coexistence of In$_2$Se$_3$ and InSe; whereas regions B and C show pure In$_2$Se$_3$ with a homogeneous distribution of indium and selenium. (c) XRD patterns of the crystals from regions B and C show only (00l), indicative of their single crystalline nature.

Additionally, we performed electronic transport measurements on Batch-2 $\alpha-In_2Se_3$ single crystal using PPMS to assess the density of Se-vacancies. The in-plane resistivity ($\rho_{xx}$) at 300K is about 63 mΩcm, which is only 0.3% of the $\rho_{xx}$ value observed for the Batch-1 crystal (20 Ωcm) at the same temperature. **Figure 7a** shows the variation of $\rho_{xx}$ with temperature. Batch-2 $\alpha-In_2Se_3$ single crystals displayed a broad peak in $\rho_{xx}$, followed by a metallic behavior, unlike Batch-1 $\alpha-In_2Se_3$ single crystals which showed an insulating-like behavior. The Hall resistivity ($\rho_{xy}$) as a function of magnetic field showed a negative slope **(Fig. 7b)**, confirming n-type behavior, consistent with both previous literature reports and our Batch-1 results. The room temperature carrier density evaluated from the Hall resistivity data **(Fig. 7c)** for Batch-2 is $2.32 \times 10^{17}$ cm$^{-3}$, smaller than the values reported for crystals grown by the conventional Bridgman method ($5 \times 10^{17} - 2 \times 10^{18}$ cm$^{-3}$ at 300K)[29–32] but nearly an order of magnitude higher than the Batch-1 sample. This increased carrier density of the Batch-2 crystal indicates a higher concentration of Se vacancies in $\alpha-In_2Se_3$ single crystals from Batch-2 than Batch-1, likely arising from the use of a reduced Se-flux (10% lower than what was used for Batch-1 growth).

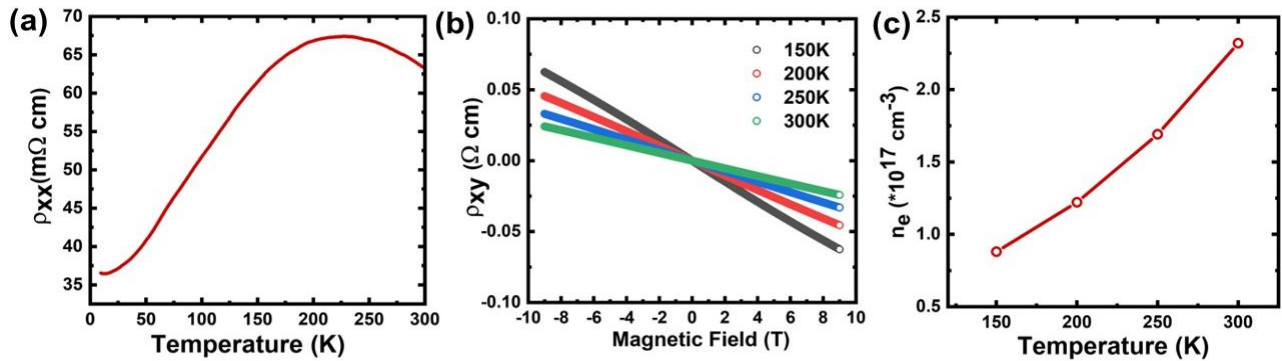

Figure 7. (a) Temperature dependence of the in-plane resistivity ($\rho_{xx}$) of Batch-2 $\alpha-In_2Se_3$ crystal. (b) Magnetic-field dependence of the Hall resistivity ($\rho_{xy}$) showing negative slope, confirming electrons to be the dominant carriers in the Batch-2 grown $\alpha-In_2Se_3$. (c) Temperature dependence of the carrier density ($n_e$) extracted from the Hall resistivity data.

**Comparison between Batch-1 and Batch-2 crystals based on Angle-resolved Photoemission Spectroscopy (ARPES)**

To further assess the electron doping in Batch-1 and Batch-2 $\alpha-In_2Se_3$ single crystals, the electronic band structure of the bulk crystals was examined by ARPES. The ARPES spectra at 300K for two Batch-1 single crystal pieces (S1 and S2) and one Batch-2 single crystal (S1) piece are presented in **Fig. 8 (top panel)** and the corresponding second derivatives are shown in **Fig. 8 (bottom panel)**. The Batch-2 sample exhibited a well-defined inverted parabolic valence band characteristic of a Mexican hat-like band dispersion[43]. It is also noted that the conduction band minima and the top of the valence band maxima are located close to the center of the Brillouin zone. In contrast, both Batch-1 crystal pieces showed a smeared density of states rather than distinct clear bands, likely due to the surface roughness that remained even after cleaving. We observed no signatures of bulk conduction near the Fermi level ($E_F$) in one of the Batch-1 samples **(Fig. 8a and 8b)**, suggesting that the sample is nearly undoped. However, The ARPES spectrum for the second Batch-1 sample (Batch1-S2) and the Batch-2 sample (Batch2-S1) showed traces of bulk conduction across $E_F$, revealing inhomogeneous doping[43], which can be attributed to Se vacancies. The $E_F$ lies approximately 0.20 eV above the conduction band minima (CBM) for Batch1-S2 and 0.31 eV above CBM in Batch2-S1.

Furthermore, the band gap values were also estimated from the ARPES spectra. Batch1-S1 and Batch1-S2 crystals have a gap of ~1.7eV whereas Batch2-S1 shows a narrower band gap ~1eV, which is consistent with our observation of insulating behavior and reduced carrier density in Batch-1 crystal compared to Batch-2. Although the Batch1 crystal was grown with higher Se flux and exhibited reduced electron density, implying fewer Se vacancies, some crystal pieces still show bulk conduction near $E_F$, thereby indicating that the compositional distribution remains inhomogeneous to some extent. This inhomogeneity remains undetected by EDXS, probably because of its limited resolution. The observed compositional inhomogeneity likely originates from the use of insufficient Se-flux during the growth and can probably be solved by further increasing the Se-flux during the growth.

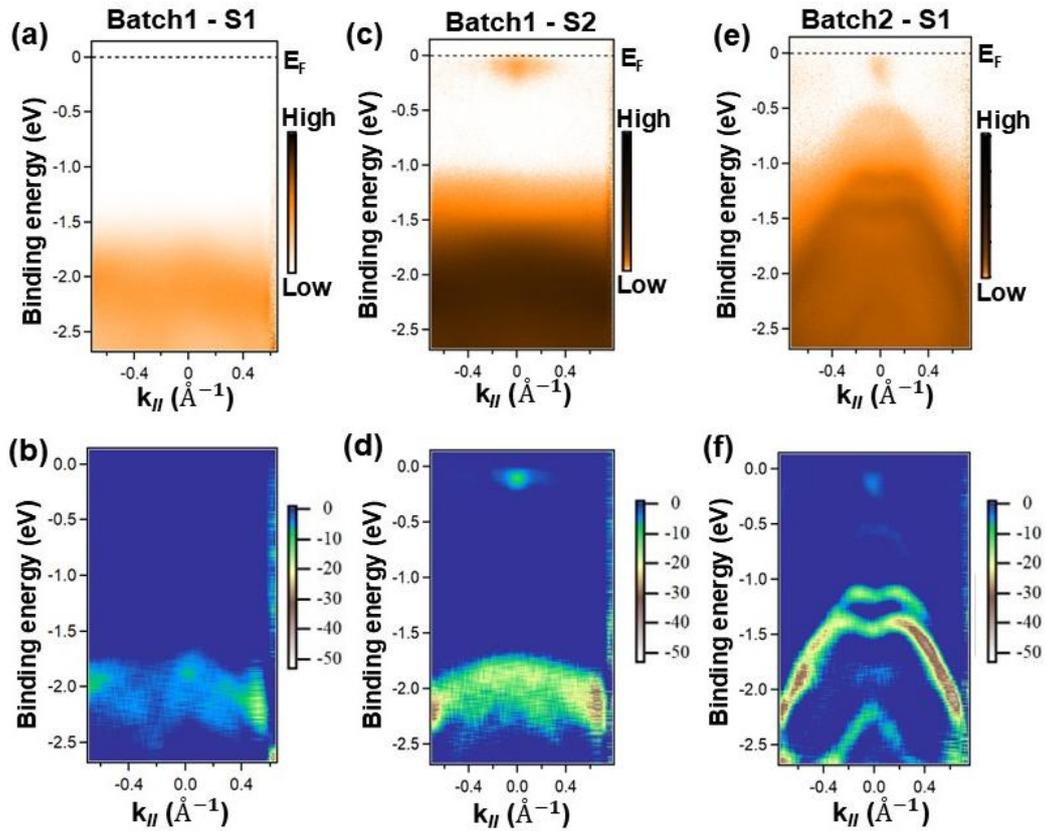

Figure 8. (a) ARPES spectra for a Batch-1 $\alpha-In_2Se_3$ crystal piece (S1) at 300K. The color bar represents counts in arbitrary units. (b) The corresponding second derivative of the Batch1-S1 ARPES spectra. (c) ARPES spectra for another Batch-1 crystal piece (S2) at 300K. (d) Second derivation of the spectra shown in (c). (e) Same as (a) and (c) for a representative Batch-2 single crystal displaying an "inverted Mexican hat-like" bands. (f) Second derivative of the Batch2-S1 ARPES spectra.

**Conclusion**

In summary, we synthesized $\alpha-In_2Se_3$ single crystals in a modified Bridgman furnace with Se-flux and liquid encapsulation under high pressure. Despite this modified Bridgman furnace having a double-crucible geometry, we could not employ the feeding method to supply source material during growth, because of high volatility of $In_2Se_3$. Instead, we grew two batches of $In_2Se_3$ using different excess Se concentrations close to the eutectic point. Structural and compositional characterization results confirmed the growth of pure $\alpha-In_2Se_3$ single crystals with 3R stacking sequence. Batch-1 crystals revealed lowest carrier density reported so far (1.5-3.2×10$^{16}$ cm$^{-3}$ at 300K), about 1-2 orders of magnitudes lower than those grown using conventional methods, indicating a significant reduction in defect concentration. Though Batch-1 crystals show an overall insulating behavior, observation of bulk conduction near $E_F$ in some pieces, together with difficulties in getting reliable P-E loop data, suggests that the carrier density needs to be further reduced. As discussed above, this can likely be achieved by increasing Se-flux concentration during growth. In contrast, Batch-2 $\alpha-In_2Se_3$ exhibited carrier density of $2.32\times10^{17}$ cm$^{-3}$ at 300K. This comparison further elucidates the critical role of Se-flux concentration in determining the stoichiometry and defect landscape of $\alpha-In_2Se_3$ crystals. A higher Se-flux (29.3% excess) in Batch-1 effectively suppressed Se vacancies, resulting in a lower carrier density and more insulating behavior, whereas reducing the Se excess to 20% in Batch-2 resulted in a metallic-like behavior and a correspondingly higher carrier concentration. These results underscore the strong sensitivity of $In_2Se_3$ crystal growth to the Se-flux ratio used and encapsulation conditions, which are crucial for achieving near stoichiometric compositions with minimal Se vacancies.

The successful growth of $In_2Se_3$ further highlights the potential of this method as a promising route for growing large single crystals, particularly for metal chalcogenides with complex phase diagram and volatile constituent elements, marking a significant advancement in materials synthesis techniques.

**Supplementary Material**

The supplementary material contains the simulated electron diffraction patterns used to compare the electron diffraction pattern of our crystal (**Fig. S1**) and the TEM images from slightly different regions of the sample revealing some extra reflections possibly originating from the presence of twin planes (**Fig. S2**).

**Acknowledgements**

The crystal growth and characterization efforts were primarily supported by the U.S. Department of Energy under grant DE-SC0019068. The transport and ARPES measurements are supported the U.S. National Science Foundation (FuSe-2425599). The TEM analysis was supported by the U.S. Department of Energy under grant DE-SC0024943. S.V.G.A. and N.A. acknowledge the support from the NSF through the Pennsylvania State University Materials Research Science and Engineering Center (MRSEC) DMR-2011839 (2020-2026).

**Author Declarations**

  **Conflict of Interest**

The authors have no conflicts to disclose**Author Contributions**

S.M. and Y.G. carried out the crystal growth of α-In$_2$Se$_3$ under the guidance of Z.M. S.M. performed XRD and EDXS to characterize the grown crystals and performed the transport measurements. S.A., S.V.G.A and N.A. performed TEM analysis on the α-In$_2$Se$_3$ single crystals. A.P. helped to perform Laue to identify crystallographic axis for TEM measurement. S.I., S.S. and N.S. did ARPES measurements on the α-In$_2$Se$_3$ single crystals. The manuscript is prepared by S.M. and Z.M. with input from all co-authors.

**Data Availability**

The data that support the findings of this study are available from the corresponding author upon reasonable request.

**References:**

[1] J. Cui, H. Peng, Z. Song, Z. Du, Y. Chao, and G. Chen, "Significantly Enhanced Thermoelectric Performance of γ-In$_2$Se$_3$ through Lithiation via Chemical Diffusion," Chem. Mater. **29**(17), 7467–7474 (2017).

[2] H. Yuan, and Y.W. Zhang, "Role of Ferroelectric In$_2$Se$_3$ in Polysulfide Shuttling and Charging/Discharging Kinetics in Lithium/Sodium-Sulfur Batteries," ACS Appl. Mater. Interfaces **14**(14), 16178–16184 (2022).

[3] G. Karmakar, D. Dutta Pathak, A. Tyagi, B.P. Mandal, A.P. Wadawale, and G. Kedarnath, "Molecular precursor mediated selective synthesis of phase pure cubic InSe and hexagonal In$_2$Se$_3$ nanostructures: new anode materials for Li-ion batteries," Dalt. Trans. **52**(20), 6700–6711 (2023).

[4] G. Almeida, S. Dogan, G. Bertoni, C. Giannini, R. Gaspari, S. Perissinotto, R. Krahne, S. Ghosh, and L. Manna, "Colloidal Monolayer β-In$_2$Se$_3$ Nanosheets with High Photoresponsivity," J. Am. Chem. Soc. **139**(8), 3005–3011 (2017).

[5] H. Lee, Y.K. Kim, D. Kim, and D.H. Kang, "Switching behavior of indium selenide-based phase-change memory cell," IEEE Trans. Magn. **41**(2), 1034–1036 (2005).

[6] S. Chen, X. Liu, X. Qiao, X. Wan, K. Shehzad, X. Zhang, Y. Xu, and X. Fan, "Facile Synthesis of γ-In$_2$Se$_3$ Nanoflowers toward High Performance Self-Powered Broadband γ-In$_2$Se$_3$/Si Heterojunction Photodiode," Small **13**(18), 1–7 (2017).

[7] R.B. Jacobs-gedrim, M. Shanmugam, N. Jain, and C. a Durcan, "Supporting information Extraordinary Photoresponse in Two-Dimensional In$_2$Se$_3$ Nanosheets," (1), 514–521 (2014).

[8] Y. Hu, W. Feng, M. Dai, H. Yang, X. Chen, G. Liu, S. Zhang, and P. Hu, "Temperature-dependent growth of few layer β-InSe and α-In$_2$Se$_3$ single crystals for optoelectronic device," Semicond. Sci. Technol. **33**(12), (2018).

[9] Y. Zhou, D. Wu, Y. Zhu, Y. Cho, Q. He, X. Yang, K. Herrera, Z. Chu, Y. Han, M.C. Downer, H. Peng, and K. Lai, "Out-of-Plane Piezoelectricity and Ferroelectricity in Layered α-In$_2$Se$_3$ Nanoflakes," Nano Lett. **17**(9), 5508–5513 (2017).

[10] C. Cui, W.J. Hu, X. Yan, C. Addiego, W. Gao, Y. Wang, Z. Wang, L. Li, Y. Cheng, P. Li, X. Zhang, H.N. Alshareef, T. Wu, W. Zhu, X. Pan, and L.J. Li, "Intercorrelated In-Plane and Out-of-Plane Ferroelectricity in Ultrathin Two-Dimensional Layered Semiconductor In$_2$Se$_3$," Nano Lett. **18**(2), 1253–1258 (2018).

[11] S. Wan, Y. Li, W. Li, X. Mao, W. Zhu, and H. Zeng, "Room-temperature ferroelectricity and a switchable diode effect in two-dimensional α-In$_2$Se$_3$ thin layers," Nanoscale **10**(31), 14885–14892 (2018).

[12] B. Lv, Z. Yan, W. Xue, R. Yang, J. Li, W. Ci, R. Pang, P. Zhou, G. Liu, Z. Liu, W. Zhu, and X. Xu, "Layer-dependent ferroelectricity in 2H-stacked few-layer α-In$_2$Se$_3$," Mater. Horizons **8**(5), 1472–1480 (2021).

[13] L. Bai, C. Ke, Z. Luo, T. Zhu, L. You, and S. Liu, "Intrinsic Ferroelectric Switching in Two-Dimensional α-In$_2$Se$_3$," ACS Nano, (2024).

[14] J. Kim, S. Kim, and H. Yu, "Enhanced Electrical Polarization in van der Waals α-

In$_2$Se$_3$ Ferroelectric Semiconductor Field-Effect Transistors by Eliminating Surface Screening Charge," Small **20**(50), 1–13 (2024).

[15] W. Ding, J. Zhu, Z. Wang, Y. Gao, D. Xiao, Y. Gu, Z. Zhang, and W. Zhu, "Prediction of intrinsic two-dimensional ferroelectrics in In$_2$Se$_3$ and other III2-VI3 van der Waals materials," Nat. Commun. **8**(1), 14956 (2017).

[16] S.M. Nahid, S. Nam, and A.M. van der Zande, "Depolarization Field-Induced Photovoltaic Effect in Graphene/α-In$_2$Se$_3$/Graphene Heterostructures," ACS Nano **18**(22), 14198–14206 (2024).

[17] H. Wang, S. Wu, Y. Chen, Q. Zhao, J. Zeng, R. Yin, Y. Zheng, C. Liu, S. Zhang, T. Lin, H. Shen, X. Meng, J. Ge, X. Wang, J. Chu, and J. Wang, "Bulk photovoltaic effect in two-dimensional ferroelectric α-In$_2$Se$_3$," Sci. China Inf. Sci. **68**(2), 122401 (2025).

[18] Z. Lei, J. Chang, Q. Zhao, J. Zhou, Y. Huang, Q. Xiong, and X. Xu, "Ultrafast photocurrent hysteresis in photoferroelectric α-In$_2$Se$_3$ diagnosed by terahertz emission spectroscopy," Sci. Adv. **11**(7), 1–10 (2025).

[19] M. Küpers, P.M. Konze, A. Meledin, J. Mayer, U. Englert, M. Wuttig, and R. Dronskowski, "Controlled Crystal Growth of Indium Selenide, In$_2$Se$_3$, and the Crystal Structures of α-In$_2$Se$_3$," Inorg. Chem. **57**(18), 11775–11781 (2018).

[20] L. Liu, J. Dong, J. Huang, A. Nie, K. Zhai, J. Xiang, B. Wang, F. Wen, C. Mu, Z. Zhao, Y. Gong, Y. Tian, and Z. Liu, "Atomically Resolving Polymorphs and Crystal Structures of In$_2$Se$_3$," Chem. Mater. **31**(24), 10143–10149 (2019).

[21] C. Manolikas, "New results on the phase transformations of In$_2$Se$_3$," J. Solid State Chem. **74**(2), 319–328 (1988).

[22] J.Y. Jiping Ye, S.S. Sigeo Soeda, Y.N. Yoshio Nakamura, and O.N. Osamu Nittono, "Crystal Structures and Phase Transformation in In$_2$Se$_3$ Compound Semiconductor," Jpn. J. Appl. Phys. **37**(8R), 4264 (1998).

[23] G. Han, Z. Chen, J. Drennan, and J. Zou, "Indium Selenides: Structural Characteristics, Synthesis and Their Thermoelectric Performances," Small **10**(14), 2747–2765 (2014).

[24] K. Kambas, J. Fotsing, E. Hatzikraniotis, and C. Julien, "Characteristics of disorder in defective layered semiconductor α-In$_2$Se$_3$," Phys. Scr. **37**(3), 397–400 (1988).

[25] J. Cui, L. Wang, Z. Du, P. Ying, and Y. Deng, "High thermoelectric performance of a defect in α-In$_2$Se$_3$-based solid solution upon substitution of Zn for In," J. Mater. Chem. C **3**(35), 9069–9075 (2015).

[26] K. Kambas, and J. Spyridelis, "Far infrared optical study of α-In$_2$Se$_3$ compound," Mater. Res. Bull. **13**(7), 653–660 (1978).

[27] H.M. Oh, H. Park, C.D. Kim, Y.S. Kim, K.W. Jang, and W.T. Kim, "Growth and Characterization of α-In$_2$Se$_3$ Single Crystals Grown by Using the Chemical Transport Reaction Technique," J. Korean Phys. Soc. **43**(2), 312–314 (2003).

[28] C. De Blasi, A.V. Drigo, G. Micocci, A. Tepore, and A.M. Mancini, "Preparation and characterization of In$_2$Se$_3$ crystals," J. Cryst. Growth **94**(2), 455–458 (1989).

[29] N.O. Kim, H.G. Kim, H.J. Lim, C. Il Lee, M.S. Jin, C.S. Yoon, and W.T. Kim, "Electrical and optical properties of α-In$_2$Se$_3$ single crystals with an indium excess," J. Korean Phys. Soc. **38**(4), 405–408 (2001).

[30] N.D. Raranskii, V.N. Balazyuk, Z.D. Kovalyuk, N.I. Mel'nik, and V.B. Gevik, "Crystal growth and elastic properties of In$_2$Se$_3$," Inorg. Mater. **47**(11), 1174–1177 (2011).

[31] Y.I. Zhirko, V.M. Grekhov, and Z.D. Kovalyuk, "Characterization, optical properties and electron (exciton)-phonon interaction in bulk In$_2$Se$_3$ crystals and InSe nanocrystals in In$_2$Se$_3$ confinement," J Nanomed Nanosci **80**, 148 (2018).

[32] T.H. Nguyen, V.Q. Nguyen, A.T. Duong, and S. Cho, "2D semiconducting α-In$_2$Se$_3$ single crystals: Growth and huge anisotropy during transport," J. Alloys Compd. **810**, 151968 (2019).

[33] G. Micocci, A. Tepore, R. Rella, and P. Siciliano, "Electrical Characterization of In$_2$Se$_3$ Single Crystals," Phys. Status Solidi **126**(2), 437–442 (1991).

[34] C. Julien, M. Eddrief, M. Balkanski, E. Hatzikraniotis, and K. Kambas, "Electrical transport properties of In$_2$Se$_3$," Phys. Status Solidi **88**(2), 687–695 (1985).

[35] C. Julien, E. Hatzikraniotis, and K. Kambas, "Electrical transport properties of impurity-doped In$_2$Se$_3$," Phys. Status Solidi **97**(2), 579–585 (1986).

[36] C. Julien, and M. Balkanski, "Optical and electrical transport studies in Li$_{0.1}$In$_2$Se$_3$," Mater. Sci. Eng. B **38**(1–2), 1–8 (1996).

[37] Y. Dai, S. Zhao, H. Han, Y. Yan, W. Liu, H. Zhu, L. Li, X. Tang, Y. Li, H. Li, and C. Zhang, "Controlled Growth of Indium Selenides by High-Pressure and High-Temperature Method," Front. Mater. **8**(January), 1–7 (2022).

[38] C. Tang, Y. Sato, K. Watanabe, T. Tanabe, and Y. Oyama, "Selective crystal growth of indium selenide compounds from saturated solutions grown in a selenium vapor," Results Mater. **13**, 100253 (2022).

[39] Y. Guan, S. Yoshida, J. Obando-Guevara, S. Mondal, S.H. Lee, H. Pfau, and Z. Mao, "Double-Crucible Vertical Bridgman Technique for Stoichiometry-Controlled Chalcogenide Crystal Growth," Cryst. Growth Des., (2025).

[40] H.-G. Kim, I. Min, and W. Kim, "Phase transition temperature of the α-In$_2$Se$_3$ single crystal," Solid State Commun. **64**(5), 819–822 (1987).

[41] H. Okamoto, "In-Se (indium-selenium)," J. Phase Equilibria Diffus. **25**(2), 201–201 (2004).

[42] H. Bergeron, L.M. Guiney, M.E. Beck, C. Zhang, V.K. Sangwan, C.G. Torres-Castanedo, J.T. Gish, R. Rao, D.R. Austin, S. Guo, D. Lam, K. Su, P.T. Brown, N.R. Glavin, B. Maruyama, M.J. Bedzyk, V.P. Dravid, and M.C. Hersam, "Large-area optoelectronic-grade InSe thin films via controlled phase evolution," Appl. Phys. Rev. **7**(4), (2020).

[43] G. Kremer, A. Mahmoudi, M. Bouaziz, M. Rahimi, F. Bertran, J. Dayen, M.L. Della Rocca, M. Pala, A. Naitabdi, J. Chaste, F. Oehler, and A. Ouerghi, "Mexican hat–like valence band dispersion and quantum confinement in rhombohedral ferroelectric alpha-In$_2$Se$_3$," Phys. Rev. B **112**(15), 155418 (2025).

[44] H. Ji, A. Reijnders, T. Liang, L.M. Schoop, K.S. Burch, N.P. Ong, and R.J. Cava, "Crystal structure and elementary electronic properties of Bi-stabilized α-$In_2Se_3$," Mater. Res. Bull. **48**(7), 2517–2521 (2013).

# Successful growth of low carrier density α-In$_2$Se$_3$ single crystals using Se-flux in a modified Bridgman furnace

Soumi Mondal[1], Sreekant Anil[2], Saurav Islam[1], Yingdong Guan[1], Sai Venkata Gayathri Ayyagari[2], Aaron Pearre[1], Sandra Santhosh[1], Nasim Alem[2], Nitin Samarth[1], Zhiqiang Mao[1,2a]

[1]Department of Physics, The Pennsylvania State University, University Park, Pennsylvania 16802, USA

[2]Department of Materials Science and Engineering, The Pennsylvania State University, University Park, PA 16802, USA

[a] Author to whom correspondence should be addressed: zim1@psu.edu

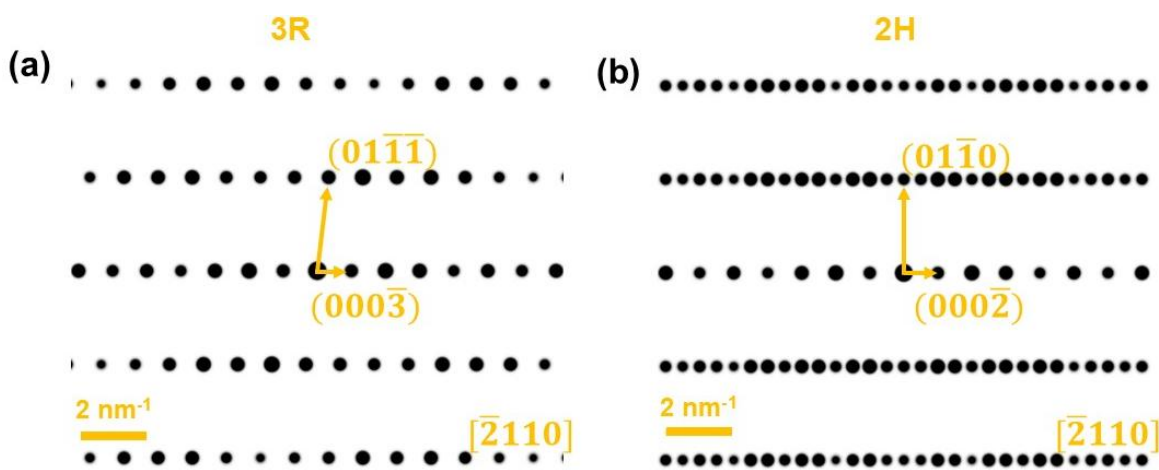

Figure S1. Simulated electron diffraction patterns along [$\bar{2}$110] zone axis of (a) 3R-α-In$_2$Se$_3$ and (b) 2H-α-In$_2$Se$_3$. The CIF file used was obtained from CCDC Deposition Number 1855522. The cell parameters for the 2H phase were taken from the supplementary information of the paper: **'Controlled Crystal Growth of Indium Selenide, In2Se3, and the Crystal Structures of α-In2Se3'**[1]. The diffraction patterns were simulated using SingleCrystal software

The simulated electron diffraction patterns along the [$\bar{2}$110] zone axis for 3R α−In$_2$Se$_3$ and 2H α−In$_2$Se$_3$ are shown in **Fig. S1a** and **Fig. S1b** respectively. Although the sample adopts the 3R α-In$_2$Se$_3$ structure, it contains twin planes. The diffraction pattern in **Fig. S2b** exhibits extra reflections and streaks, which arise from the presence of a twin plane (indicated in **Fig. S2a, c**). Based on the symmetry of the pattern, these twin planes are related by a mirror plane oriented parallel to the c-axis. The ADF-STEM image (**Fig. S2c**) further confirms this, as the atomic arrangement on either side of the white line is mirrored with respect to a plane parallel to the c-axis.

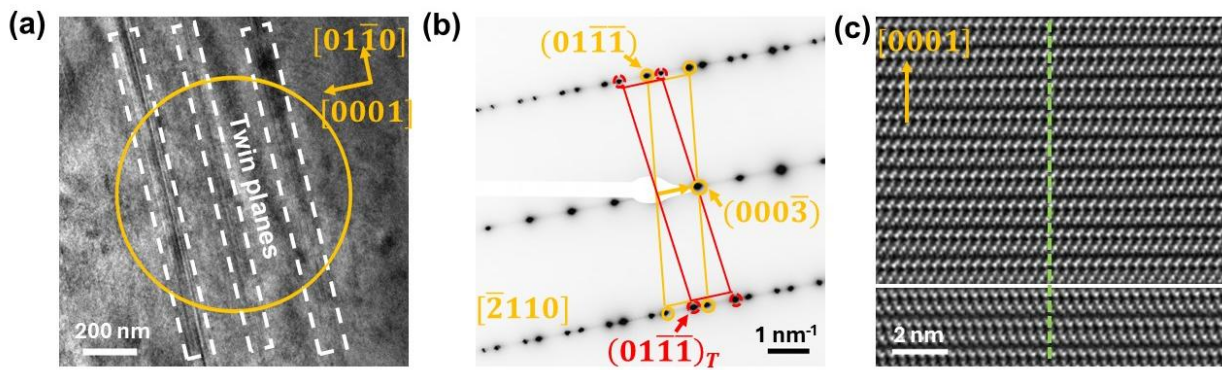

Figure S2. (a) TEM image of 3R-α-In$_2$Se$_3$ viewed along the [$\bar{2}$110] zone axis. (b) Corresponding SAED pattern from the circled region in (a), displaying additional reflections associated with the twin plane. (c) ADF-STEM image in which the twin boundary is clearly visible, indicated by the white line.

References:

[1] M. Küpers, P.M. Konze, A. Meledin, J. Mayer, U. Englert, M. Wuttig, and R. Dronskowski, "Controlled Crystal Growth of Indium Selenide, In$_2$Se$_3$ , and the Crystal Structures of α-In$_2$Se$_3$," Inorg. Chem. **57**(18), 11775–11781 (2018).